\documentstyle[12pt]{article}
\begin{document}
\begin{flushright}
hep-th/0303071\\
SNBNCBS-2003
\end{flushright}
\vskip 3.5cm
\begin{center}
{\bf \Large { Lagrangian formulation of some $q$-deformed systems}}

\vskip 3.5cm

{\bf R.P.Malik}
\footnote{ E-mail address: malik@boson.bose.res.in  }
\footnote{ Present address: S. N. Bose National Centre for Basic Sciences,
Block-JD, Sector-III, Salt Lake, Calcutta--700 098, West Bengal, India.}\\
{\it Bogoliubov Laboratory of Theoretical Physics,} \\
{\it Joint Institute for Nuclear Research,
P. O. Box 79, 101000 Moscow, Russia}

\vskip 3.5cm

\end{center}

\noindent
{\bf Abstract}:
In the framework of Lagrangian formulation, some $q$-deformed physical
systems are considered. The $q$-deformed Legendre transformation is 
obtained for the free motion of a non-relativistic particle on a quantum line. 
This is subsequently exploited to obtain the 
Lagrangians for the $q$-deformed harmonic
oscillator and $q$-deformed relativistic free particle. The Euler-Lagrange
equations of motion are derived in a consistent manner with the
corresponding Hamilton's equations. The Lagrangian for the $q$-deformed
relativistic particle is endowed with the $q$-deformed gauge symmetry and
reparametrization invariance which are shown to be equivalent only
for $ q = \pm 1$.\\

\baselineskip=16pt


\newpage

\noindent
The concept of deformation has played a notable part in the development of
modern theoretical physics. The most familiar and cardinal examples of such
a class of deformed physical theories are presumed to be the quantum mechanics 
and the special theory of relativity with deformation parameters like
Planck's constant $(\hbar)$ and the speed of light $(c)$, respectively [1,2]. 
A key feature of
thorough understanding of these theories is the emergence of a couple of
fundamental constants of nature, namely; $\hbar$ and $c$. It is conjectured 
that the deformation of groups, based on the quasi-triangular Hopf algebras
[3,4], together with the ideas of non-commutative geometry might provide
a ``fundamental length'' in the context of spacetime quantization [5] which
would have close kinship with the dimensionless deformation parameter
$(q)$ of the deformed groups [6]. Some attempts have recently been made
to associate ``$q$'' with the relativistic quantities [7] and the length of
compactification [8] in the context of some concrete physical examples. In 
addition, these deformed (so-called quantum) groups have also been treated
as gauge groups for the development of the $q$-deformed gauge theories [9].

It is an interesting endeavour to apply the ideas of the quantum groups in
a cogent way to some known physical systems [7-10]. The purpose of the
present paper is to develop the Lagrangian formulation for some known
physical systems by exploiting the basic ingredients of the quantum group
$GL_{q} (2)$ and the corresponding differential calculus [11] discussed
on the quantum phase space [12]. We obtain the $q$-deformed Legendre
transformations and relevant Euler-Lagrange equations of motion for
a free non-relativistic particle, a harmonic oscillator and a relativistic
particle on a quantum line which are consistent with the $q$-deformed
Hamilton's equations of motion. One of the salient features of our approach
is that the equations of motion for a given $q$-deformed physical system
remain {\it the same} as that of its undeformed (classical) counterpart
but the momentum, velocity and force, etc., depend on the deformation
parameter $q$. It is fascinating to find that the mass and metric in the
case of the $q$-deformed relativistic particle turn out to be
non-commutative objects on the quantum world-line, embedded in a
$D$-dimensional undeformed flat Minkowski space.

We start off with the free motion of a non-relativistic particle on a
quantum-line [12] characterized by the coordinate generator $x (t)$ and
momentum generator $p (t)$ that satisfy the following relationship on
this line
\footnote{Note that the definition of the quantum-line present in Ref. [12]
would be obtained from (1) by the replacement: $q \rightarrow q^{-1}$.}

$$
\begin{array}{lcl}
x (t)\; p(t) = q\; p(t)\; x(t),
\end{array} \eqno(1)
$$
where the $q$-trajectory of the particle, moving on a $q$-deformed
cotangent manifold, is parametrized by a real commuting variable $t$. It is 
straightforward to check that the above relation is form-invariant under
the following $GL_{q} (2)$ transformations
$$
\begin{array}{lcl}
\left (\begin{array} {c} x \\ p \\
\end{array} \right )
\;\;\rightarrow \;\;
\left (\begin{array} {cc}
a & b \\ c  & d\\
\end{array} \right )\;\;
\left (\begin{array} {c}
x \\ p \\
\end{array} \right ), 
\end{array}\eqno(2)
$$
if we assume the commutativity of the phase variables with the elements
$a, b, c$ and $d$ of the $2 \times 2$ $GL_{q} (2)$ matrix obeying
the following braiding relations in rows and columns
$$
\begin{array}{lcl}
a b &=& q b a, \quad c d = q d c, \quad ac = q ca, \quad b c = c b,
\nonumber\\
b d &=& q d b, \qquad a d - d a = (q - q^{-1})\; bc. 
\end{array} \eqno(3)
$$
To develop the Lagrangian formulation for a given classical system, it
is essential to discuss its dynamics in the tangent (velocity phase)
space. The  second-order Lagrangian $(L_{s})$ describing the free
motion $(m \ddot x = 0)$ in this space is
$$
\begin{array}{lcl}
L_{s} = {\displaystyle \frac {q} { 1 + q^2}}\; m \dot x^2,
\end{array} \eqno(4)
$$
where, in addition to $ x (t)$ and $ p (t)$, the $t$-independent
mass parameter $m$, is also a Hermitian element of an
algebra in involution (i.e. $ |q| = 1$) and $\dot x = d x/ dt$.

The most basic geometrical object in classical mechanics is the
non-degenerate and closed two-form symplectic structure, defined on
a symplectic (cotangent) manifold. The covariant and contravariant
sympletic metrics that reduce to their classical canonical counterparts
in the limit $q \rightarrow 1$, are (see, e.g., Ref. [2])
$$
\begin{array}{lcl}
\Omega^{AB} (q) =
\left (\begin{array} {cc}
0 & - q^{-1/2}\\
q^{1/2} & 0\\
\end{array} \right ),
\qquad
\Omega_{AB} (q) =
\left (\begin{array} {cc}
0 & q^{-1/2} \\
- q^{1/2} & 0 \\
\end{array} \right ).
\end{array} \eqno(5)
$$
The first-order Lagrangian ($L_{f}$) describing the free motion
($\dot p = 0$) can be obtained by exploiting the covariant metric
$\Omega_{AB} (q)$ in the following Legendre transformations
$$
\begin{array}{lcl}
L_{f} = z^A\; \bar \Omega_{AB} (q)\; \dot z^B - H = q^{1/2} p \dot x - H,
\end{array}\eqno(6)
$$
where $z^A = (x, p), H$ is the Hamiltonian function defined on the cotangent
manifold and the general expression (see, e.g., Ref. [8])
$\bar \Omega_{AB} (q) = \int_{0}^{1} \Omega_{AB} (\alpha z) \alpha d \alpha$
reduces in our case to $\bar \Omega_{AB} (q) = \frac{1}{2}\; \Omega_{AB} (q)$.
The definition of the canonical momentum $(p)$ crucially depends upon the
choice of the symplectic metric and the (non-)commutativity of the velocity
$(\dot x)$ and momentum $(p)$ in the Legendre transformations (6). For the
first- and second-order Lagrangians, the consistent expression for this 
quantity is
$$
\begin{array}{lcl}
p = q^{-3/2}\; \Bigl ( {\displaystyle \frac{\partial L_{(f,s)}} {\partial
\dot x}} \Bigr ) \equiv q^{1/2}\; m \dot x,
\end{array}\eqno(7)
$$
where the on-shell non-commutative relations $\dot x m = q m \dot x$ and
$ \dot x p = q p \dot x$, emerging from the $GL_{q} (2)$ invariant 
quantum-line (1), have been used in the derivation of (7). Furthermore,
consistent with these non-commutative relations, the following rule of the
differentiation has been invoked
$$
\begin{array}{lcl}
{\displaystyle \frac{\partial (y^r \dot x^s)} {\partial \dot x}}
= y^r \dot x^{s - 1} q^r {\displaystyle \frac{ 1 - q^{2 s}}
{1 - q^2}},
\end{array}\eqno(8)
$$
where $ y = m, p$ and $ r, s \in {\cal Z}$ are whole numbers but not fractions.
The Hamiltonian function $H$, describing the motion in the cotangent space,
can be obtained by the Legendre transformation ($ H = q^{1/2} p \dot x 
- L_{s}$) and equation (7). This is expressed in terms of the non-commutative
mass parameter $(m)$ and momentum $(p)$ as [12]
$$
\begin{array}{lcl}
H = {\displaystyle \frac {q^2} { 1 + q^2}}\; p\; m^{-1} \; p.
\end{array} \eqno(9)
$$
The contravariant sympletic metric of (5) is used in the computation of
the $q$-deformed Poisson brackets present in the Hamilton's equations of
motion. For instance, $\dot x = \{ x , H \}_{q} = \Omega^{AB} 
\partial_{A} x \;\partial_{B} H \equiv q^{-1/2} m^{-1} p$ and 
$\dot p = \{ p, H \}_{q} = 0$
are obtained due to the $GL_{q} (2)$ invariant differential calculus
defined on the phase space. All the on-shell, associative and non-commutative
relations, resulting from defining quantum-line equation (1), are listed below
$$
\begin{array}{lcl}
\dot x \; p &=& q\; p\; \dot x, \qquad p\; m = q\; m\; p, \nonumber\\
\dot x\; m &=& q\; m\; \dot x, \quad x \; m = q\; m \;x, \quad x \;
\dot x = \dot x\; x.
\end{array}\eqno(10)
$$
In the computation of the Hamilton's equations of motion ($\dot x$ and 
$\dot p$), the Hamiltonian is first of all recast in the monomial form
$m^r\; p^s$ ($ r, s \in {\cal Z}$ are whole numbers but not fractions) and
the following differentiation rule is used
$$
\begin{array}{lcl}
{\displaystyle \frac{\partial (m^r \;p^s)} {\partial p}}
= m^r \; p^{s - 1} q^r {\displaystyle \frac{ 1 - q^{2 s}}
{1 - q^2}}.
\end{array}\eqno(11)
$$

One of the key features of our discussion is that the $q$-dependence appears
only in the expressions for velocity and momentum but the equations of motion
remain the same as that in the undeformed case. In addition to the description
of quantization on a quantum-line, it has been demonstrated in Ref. [12] that
the solutions of the equations of motion respect $GL_{q} (2)$ invariance at
any arbitrary time $t$.

The Hamilton's equations of motion can be derived by requiring the invariance
of the action ($ S = \int L_{f} dt$) in the framework of the principle of
the least action, as illustrated below
$$
\begin{array}{lcl} 
\delta S = 0 = {\displaystyle \int}\; \Bigl ( q^{1/2} \;\delta p\; \dot x +
q^{1/2}\; p \; \delta \dot x - \delta x\; {\displaystyle \frac{\partial H}
{\partial x} - \delta p \; \frac{\partial H} {\partial p}} \Bigr )\; dt,
\end{array}\eqno(12)
$$
where Hamiltonian is assumed to possess no explicit time dependence. Now
taking all the variations to the left side by exploiting the following
on-shell $q$-commutation relations resulting from (1)
$$
\begin{array}{lcl}
\delta \dot x\; p = q\; p\; \delta \dot x, \qquad
\dot x\; \delta p = q\; \delta p \; \dot x,
\end{array}\eqno(13)
$$
and dropping off the total derivative term by choosing appropriate boundary
conditions on the transformation parameters, we obtain the following equations
of motion
$$
\begin{array}{lcl}
\dot x = q^{-1/2}\; {\displaystyle \frac{\partial H} {\partial p}} \qquad
\dot p = - q^{1/2}\; {\displaystyle \frac{\partial H} {\partial x}},
\end{array}\eqno(14)
$$
which are in total agreement with the choice of the contravariant metric (5)
and the $q$-deformed Poisson-brackets.

To derive the deformed Euler-Lagrange equations of motion, it is instructive
to consider the $q$-deformed harmonic oscillator on the quantum-line (1). The
Hamiltonian, first- and second-order Lagrangians for this system are
$$
\begin{array}{lcl}
H^{osc} &=& 
{\displaystyle \frac{q^2} {1 + q^2}\; p\; m^{-1} \;p
+ \frac{q^{-2} \omega^2}{1 + q^2}\; x\;m\; x}, \nonumber\\
L_{f}^{osc} &=& q^{1/2}\; p \;\dot x -
{\displaystyle \frac{q^2} {1 + q^2}\; p\; m^{-1} \;p
- \frac{q^{-2} \omega^2}{1 + q^2}\; x\;m\; x}, \nonumber\\
L_{s}^{osc} &=&
{\displaystyle \frac{q} {1 + q^2}\;  m \; \dot x^2
- \frac{q^{-2} \omega^2}{1 + q^2}\; x\;m\; x}, 
\end{array}\eqno(15)
$$
where the frequency $\omega$ is a commuting number. All the $q$-commutation 
relations (10) are valid in this case as well, because, the Hamilton's 
equations of motion $\dot x = \{ x , H^{osc} \}_{q} = q^{-1/2} m^{-1} p $ and
$\dot p = \{ p, H^{osc} \}_{q} = - \omega^2 q^{1/2} m x$ do not spoil these
relations. Moreover, the extra $q$-commutation relations $x \dot p = q \dot p
x$ and $ p \dot p = \dot p p$ are automatically satisfied due to (10). The
expression for the canonical momentum $(p)$ is same as (7) and, consistent
with the Hamilton's equations, the Euler-Lagrange equations of motion
($ \ddot x = - \omega^2 x$) is
$$
\begin{array}{lcl}
 q^{-3/2}\; {\displaystyle \frac{d} {d t}}\;
\Bigl ( {\displaystyle \frac{\partial L^{osc}_{(f,s)}} {\partial
\dot x}} \Bigr ) = q^{1/2}\;
\Bigl ( {\displaystyle \frac{\partial L^{osc}_{(f,s)}} {\partial
x}} \Bigr ). 
\end{array}\eqno(16)
$$

The Hamiltonian $(H_{v}$), describing the motion of a classical $q$-deformed
particle moving under the influence of a potential $V (x)$ is as follows
$$
\begin{array}{lcl}
H_{v} = 
{\displaystyle \frac{q^2} {1 + q^2}\; p\; m^{-1} \;p}
+ \; V (x).
\end{array}\eqno(17)
$$
The first- and second-order Lagrangians can be defined in an analogous manner
as in (15). All the non-commutative relations listed above and the
Euler-Lagrange equations of motion remain the same if the potential obeys
$$
\begin{array}{lcl}
x\; {\displaystyle \frac{\partial V} {\partial x}}
 = q\;  {\displaystyle \frac{\partial V} {\partial x}} \; x.
\end{array}\eqno(18)
$$
This requirement implies that the force on the system is a non-commutative
 object. Equation (18) is satisfied by the harmonic oscillator potential
due to (10). The $GL_{q} (2)$ invariant evolution equations, quantization
and oscillator realizations have been discussed in Ref. [12].

With the above background, we shall dwell a bit more on the free motion of
a $q$-deformed relativistic particle on a quantum world-line parametrized by
a commuting evolution parameter $\tau$. A quantum world-line, traced out by
the free motion of a $q$-relativistic particle in a $D$-dimensional flat
Minkowski phase space, is defined in terms of the coordinate generator
$x_\mu$ and the momentum generator $p_\mu$ as
$$
\begin{array}{lcl}
x_\mu  (\tau)\; p^\mu (\tau) = q\; p^\mu (\tau)\; x_\mu (\tau).
\end{array} \eqno(19)
$$
It can be readily seen that the above definition of the quantum world-line
remains invariant under the following transformations
$$
\begin{array}{lcl}
x_{\mu} \rightarrow a\; x_\mu + b\; p_\mu, \qquad
p_{\mu} \rightarrow c\; x_\mu + d\; p_\mu, 
\end{array}\eqno(20)
$$
if we assume the commutativity of the phase variables with the elements
$a, b, c$ and $d$ of a $2 \times 2$ $GL_{q} (2)$ matrix obeying the braiding
relations (3). In the definition of the $q$-world line (19), the repeated
indices are summed over ($\mu = 0, 1, 2.......D-1)$ and $GL_{q} (2)$ symmetry
transformations (20) are implied for each component-pairs of the phase
variables: $(x_{0}, p_{0}), (x_{1}, p_{1}),.....(x_{D-1}, p_{D-1})$.

The Hamiltonian, describing the free motion of the $q$-relativistic 
particle, is as follows
$$
\begin{array}{lcl}
{\cal H} = 
{\displaystyle \frac{q} {1 + q^2}\; (p_\mu\; e \;p^\mu - m\; e\; m)},
\end{array}\eqno(21)
$$
where $e$ is the einbein (metric) and all the variables, except the mass 
parameter $m$, are a function of $\tau$. The Hamilton's equations of motion,
derived from the generalized form of the Poisson-bracket and symplectic metric
(5), are as follows
$$
\begin{array}{lcl}
\dot x_\mu = \{ x_{\mu}, {\cal H} \}_{q} = q^{1/2}\; e\; p_\mu, \qquad
\dot p_\mu = \{ p_\mu, {\cal H} \}_{q} = 0,
\end{array}\eqno(22)
$$
where $\dot  x_\mu = (d x_\mu / d \tau)$. The defining quantum world-line 
equation (19), together with (22) and associativity requirements, leads to
the validity of the following on-shell non-commutative relations
$$
\begin{array}{lcl}
\dot x_\mu p^\mu &=& q\; p_\mu \dot x^\mu, \quad \ddot x_\mu p^\mu = q \;p^\mu
\ddot x_\mu, \quad e m = q\; m e, \quad p_\mu m = m p_\mu, \nonumber\\
\dot x_\mu m &=& q \;m \dot x_\mu, \quad e p_\mu = q \;p_\mu e, \quad x_\mu m =
q\; m x_\mu, \quad e \dot x_\mu = q \;\dot x_\mu e.
\end{array}\eqno(23)
$$
It can be seen that these relations are consistent 
\footnote{ It should be emphasized here that the relations
$x_\mu x_\nu = x_\nu x_\mu, p_\mu p_\nu = p_\nu p_\mu, x_\mu p_\nu 
= q p_\nu x_\mu$ have played a very significant role in the construction
of a consistent dynamics in the non-commutative phase space
({\it R. P. Malik, A. K. Mishra, G. Rajasekaran, Int. J. Mod. Phys.
A 13 (1998) 4759}, arXive: hep-th/9707004). In a recent work
({\it R. P. Malik, Dynamics in a noncommutative space},
arXive: hep-th/0302224), these relations have been exploited to establish
a neat connection between some of the basic ideas of non-commutative geometry 
and a few key properties (and related notions) of the quantum groups.}
with $x_\mu x_\nu 
= x_\nu x_\mu, p_\mu p_\nu = p_\nu p_\mu, x_\mu p_\nu = q p_\nu x_\mu$ if we
assume $e x_\mu = q x_\mu e$ and use the on-shell conditions (22). The first-
and second-order Lagrangians can be derived from the Hamiltonian (21), as 
listed below
$$
\begin{array}{lcl}
L_{F} &=& q^{1/2}\; p_\mu \dot x^\mu
- {\displaystyle \frac{q} {1 + q^2}\; (p_\mu\; e \;p^\mu - m\; e \;m)},
\nonumber\\
L_{S} &=& 
 {\displaystyle \frac{q^2} {1 + q^2}\;  e^{-1}\; (\dot x_\mu)^2} 
+ {\displaystyle \frac{q} {1 + q^2}\; m\; e\; m}.
\end{array}\eqno(24)
$$
Analogous to eq. (7) and consistent with (22), the expression for the 
canonical momentum $(p_\mu)$ is
$$
\begin{array}{lcl}
p_\mu = q^{-3/2}\; \Bigl ( {\displaystyle \frac{\partial L_{(F,S)}} {\partial
\dot x}} \Bigr ) \equiv q^{-1/2}\; e^{-1}\; \dot x_\mu.
\end{array}\eqno(25)
$$
The $q$-differentiation [11] of the second-order Lagrangian $L_{S}$ with 
respect to the multiplier field ``$e$'' yields
$$
\begin{array}{lcl}
{\displaystyle \frac{q^4} {1 + q^2} \bigl (m^2 - e^{-1} \dot x_\mu e^{-1}
\dot x^\mu \bigr )} = 0,
\end{array}\eqno(26)
$$
which, in turn, leads to the mass-shell condition for the $q$-deformed free
particle as follows
$$
\begin{array}{lcl}
p_\mu\; p^\mu - m^2 = 0.
\end{array}\eqno(27)
$$
This equation is one of the Casimir invariants of the Poincar{\'e} group
corresponding to the 
undeformed Minkowski space. The eigen value of this operator
and Pauli-Lubanski vector would designate the eigen states, that would be 
needed for the representation theory of the Poincar{\' e} group. This
constraint condition is in neat conformity with the recent discussion [13]
of the Klein-Gordon equation and the Dirac equation derived from the
$q$-deformation of the Dirac $\gamma$-matrices. Furthermore, eq. (26) yields
the following relationship among einbein, velocity and mass parameter
$$
\begin{array}{lcl}
e^{-2} = m^2\; \bigl (\dot x_\mu \dot x^\mu \bigr )^{-1}.
\end{array}\eqno(28)
$$
The computation of $e$ and $e^{-1}$ from (28) is a bit tricky because of the
non-commutativity of velocity and mass. A nice and simple way to compute these
is firstly start with
$$
\begin{array}{lcl}
e^{-1} = f (q)\;m\; \bigl (\dot x_\mu \dot x^\mu \bigr )^{-1/2},
\end{array}\eqno(29)
$$
and require the validity of (28). Using the $q$-commutation relations (23),
the second-order Lagrangian $L_{S}$ can be recast in various forms where
$e^{-1}$ and $e$ would occupy different positions in its first and second
terms. The requirement of equality of the resulting Lagrangians leads to
the determination of $f (q)$ to be $q^{1/2}$ if the substitution (29) is made
\footnote{ We have also used $\dot x_\mu (\dot x)^{1/2} = (\dot x)^{1/2}
\dot x_\mu $ which results in from (29) with $ e \dot x_\mu = q \dot x_\mu e$
and $ x_\mu m = q m \dot x_\mu$.}. Ultimately, the following $q$-deformed
Lagrangian with square root is obtained from the second order Lagrangian 
$L_{S}$
$$
\begin{array}{lcl}
L_{0} = q^{1/2}\; m \; (\dot x_\mu \dot x^\mu)^{1/2}.
\end{array}\eqno(30)
$$
The action $A = q^{1/2}\; m \int_{\tau_1}^{\tau_2}\; d \tau (\dot x^2)^{1/2}
\equiv q^{1/2} m \int_{1}^{2} ds$ corresponding to (30) and, proportional
to the path length $ ds = (dx_\mu d x^\mu)^{1/2}$, is invariant under the
undeformed reparametrization transformations $ \tau \rightarrow f (\tau)$,
where $f (\tau)$ is a monotonically varying function of $\tau$. The definition
of the canonical momentum (25) is correct in the case of the above
Lagrangian too. This can be seen (with $ (\dot x^2)^{1/2} m 
= q m (\dot x^2)^{1/2}$) as follows
$$
\begin{array}{lcl}
p_\mu = q^{-3/2}\; {\displaystyle \frac{ \partial (\dot x)^2} {\partial
\dot x^\mu}\;\frac{ \partial (\dot x^2)^{1/2}} {\partial \dot x^2} 
\frac{\partial
[q^{1/2} m (\dot x_\mu \dot x^\mu)^{1/2}]}{\partial (\dot x^2)^{1/2}}}  
\equiv \dot x_\mu (\dot x^2)^{-1/2} m.
\end{array}\eqno(31)
$$
It should be emphasized here that, while computing the $q$-derivative of the
$q$-variables with fractional power, the following rule has to be invoked
$$
\begin{array}{lcl}
{\displaystyle \frac{\partial} {\partial x}} \bigl (x^{r/s} \bigr )
= {\displaystyle \frac{1 - q^{2r}} {1 - q^{2s}}}\; x^{(r/s) - 1},
\end{array}\eqno(32)
$$
where $r$ is {\it not} divisible by $s$ ($ r, s \in {\cal Z}$). Furthermore,
the mass-shell condition (27) is satisfied for both the left chain rule as 
well as the right chain rule of differentiation, implemented in the
computation of (31).

The equation of motion $ \dot p_\mu = 0$, resulting from (31), leads to the
following expression
$$
\begin{array}{lcl}
\ddot x_\mu (\dot x)^2 - \dot x_\mu (\dot x_\nu \dot x^\nu) = 0,
\end{array}\eqno(33)
$$
which corresponds to the equation of motion in the undeformed case. It is 
difficult to extract out the evolution equation at arbitrary $\tau$ from (33).
The viable alternative is to parametrize the evolution equation in terms
of the path length ``$s$'' [14]. In the purview of this change of
parametrization, the canonical momentum $p_\mu = m \;(d x_\mu / d s) \equiv
m \dot x_\mu (\dot x^2)^{-1/2}$, leads to the equation of motion
($ m \; (d^2 x_\mu / d s^2) = 0$). The evolution equations
$$
\begin{array}{lcl}
x_\mu (s) = x_\mu (0) + m^{-1}\; p_\mu (0)\; s, \qquad
p_\mu (s) = p_\mu (0),
\end{array}\eqno(34)
$$
respect the ``$GL_{q} (2)$-invariance'' (i.e. $x_\mu (s) p^\mu (s)
= q p^\mu (s) x_\mu (s)$) at any arbitrary value of the path-length $s$
because $ s p_\mu (0) = q p_\mu (0) s$ and $ s m = q m s$
\footnote{ These $q$-commutation relations are obtained due to (29),
$ e p_\mu = q p_\mu e, (\dot x^2)^{1/2} m = q m (\dot x^2)^{1/2}, p_\mu m
= m p_\mu$ and the commutativity property of $\tau$ with the 
phase variables.}. It is worth pointing 
out that, in contrast to the commutativity of $\tau$, the path length
$s$ is a non-commutative parameter which turns out to be handy only in the
description of the evolution equations.

It is important to pin-point here that, unlike the non-relativistic cases
where $ p m = q m p$, one obtains $ p_\mu m = m p_\mu$ in the case of
$q$-deformed relativistic particle. The correctness of these relations can 
be checked by using the on-shell $q$-commutation relations (10), (23) and
the substitution $p = q^{1/2} m \dot x, p_\mu = q^{-1/2} e^{-1} \dot x_\mu$.
In fact, the space part of $ p_\mu m = q^{-1/2} e^{-1} \dot x_\mu m$ reduces
to $ q^{-1/2} m \dot x m$ in the one-dimensional non-relativistic limit
which corresponds to $ \dot x_\mu \rightarrow \dot x, (\dot x^2)^{1/2}
\rightarrow 1$ and $ e^{-1} = q^{1/2} m (\dot x^2)^{-1/2} \rightarrow
 q^{1/2} m$. Now, due to (10), it is clear that $ m \dot x m = q m m \dot x$
which yields the non-relativistic relation $ p m = q m p$. This conclusion 
can also be drawn from the relativistic relation $p_\mu e^{-1} 
= q e^{-1} p_\mu$ because in the non-relativistic limit: $ p_\mu \rightarrow p,
e^{-1} = q^{1/2} m (\dot x^2)^{-1/2} \rightarrow q^{1/2} m$. The commutativity
of the mass parameter $m$ and momenta $p_\mu$ in the case of the 
$q$-relativistic particle is primarily due to the existence of the mass-shell
condition (27).

All three Lagrangians of (24) and (30) are equivalent and are endowed with
gauge and reparametrization symmetries. To illustrate this, we shall
concentrate on the first order Lagrangian $L_{F}$. It is obvious that the
$q$-canonical momentum ($\Pi_{e}$) with respect to the multiplier field
$e (\tau)$ is zero. Thus, $\Pi_{e} \approx 0$ is the primary constraint. The 
secondary constraint $\Pi_{e}^{(1)}$ can be obtained by requiring the 
consistency of the primary constraint under time evolution, generated by
the Hamiltonian ${\cal H}$. This is given by
$$
\begin{array}{lcl}
\Pi_{e}^{(1)} = \{ \Pi_{e}, {\cal H} \}_{q} = -
{\displaystyle \frac{q^{1/2} q^4} {1 + q^2}}\; (p^2 - m^2) \approx 0,
\end{array}\eqno(35)
$$
which amounts to the validity of the mass-shell condition. Both these
constraints are first-class in the language of Dirac and there are no tertiary
constraints. The gauge symmetry transformations, generated by these 
constraints, are as follows
$$
\begin{array}{lcl}
\delta x^\mu = q^{1/2}\; \xi\; p^\mu, \qquad \delta p^\mu = 0,\qquad
\delta e = q^2\; \dot \xi,
\end{array}\eqno(36)
$$
where $\xi$ is the non-commutative gauge transformation parameter. This can be
seen by the application of the transformations (36) and requiring the validity
of (22) on the $q$-deformed world-line (19) which yields $\xi p_\mu = q p_\mu
\xi$. As per our convention, all the symmetry transformations are firstly
taken to the left and then the substitutions (36) are made. The 
quasi-invariance of the Lagrangian is succinctly expressed as follows
$$
\begin{array}{lcl}
\delta L_{F} = {\displaystyle \frac{d}{d \tau} \Bigl (\frac{\xi (p^2  
+ q^2 m^2)} {(1 + q^2)} \Bigr )},
\end{array}\eqno(37)
$$
where the chain rule $d p^2/ d \tau = (d p^2/dp^\mu) (d p^\mu/ d \tau)
= (1 + q^2) p_\mu \dot p^\mu$ has been used. Even if we do not take the
symmetry variation to the left side in all the terms of $L_{F}$ but exploit
the non-commutativity of $\xi$, then also, we end up with the 
transformation (37).

In addition to the gauge symmetry, the first-order Lagrangian is invariant
under the following reparametrization transformations
$$
\begin{array}{lcl}
\delta_{r} \;x_\mu = \epsilon \;\dot x_\mu, \qquad \delta_{r} \;p_\mu =
\epsilon \;\dot p_\mu, \qquad \delta_{r}\; e = 
{\displaystyle \frac{d}{d \tau} \;(\epsilon e)},
\end{array}\eqno(38)
$$
emerging due to the one-dimensional diffeomorphism $\tau \rightarrow
\tau - \epsilon (\tau)$ with commuting infinitesimal transformation
parameter $\epsilon$. This is because of the fact that
$\delta_{r} x_\mu p^\mu = q p_\mu \delta_{r} x_\mu$ with 
$\dot x_\mu p^\mu = q p_\mu \dot x^\mu$ leads to $p_\mu \epsilon 
= \epsilon p_\mu$. In fact, the first-order Lagrangian undergoes the
following change under (38)
$$
\begin{array}{lcl}
\delta_{r} L_{F} = {\displaystyle \frac{d} {d \tau}} (\epsilon L_{F}).
\end{array}\eqno(39)
$$
In the usual undeformed ($ q = 1)$ case of the free relativistic particle,
the gauge (cf. (36)) and reparametrization (cf. (38)) symmetries are
equivalent on-shell with the identification $ \xi = \epsilon e$ [15].
However, in the deformed case, these are not equivalent because the
transformations of the einbein field, in spite of the above identification,
are not equal unless $ q = \pm 1$. This discrepancy might manifest at
very high energy and might turn out to have some significant implications
in the study of non-commutative geometry and spacetime structure at
very high energy scale.

The $q$-deformed relativistic particle presents a prototype example of
$q$-deformed Abelian gauge theories. In addition to the explicit derivation
of the Noether's theorem, it would be worthwhile to develop a $q$-deformed
BRST formalism to quantize this system on a quantum world-line. It seems,
there would not be any principal difficulties in the extension of our results
for the discussion of $q$-deformed spinning relativistic particle where
the ideas of the quantum group $GL_{q} (1|1)$ would play a prominent role.
Furthermore, it would be interesting to generalize the second-order 
Lagrangian $L_{S}$ to the $q$-deformed string action and discuss various
subtleties involved in it. We hope to come to these problems in future.

Fruitful and stimulating conversations with A. Isaev, R. Mir-Kasimov
and S. Shabanov are gratefully acknowledged.

\end{document}